\documentstyle[preprint,aps,eqsecnum]{revtex}
\begin{document}
\draft
\preprint{Alberta Thy-1-93,\,  hep-th/9411225}
\title{
Massive particle creation in a static 1+1 dimensional
spacetime}
\author{D.\ J. Lamb\cite{email} and A.\ Z. Capri}
\address{ Theoretical Physics Institute, \\
Department of Physics,
University of Alberta,\\
Edmonton, Alberta T6G 2J1, Canada}
\author{S.\ M. Roy}
\address{
Tata Institute of Fundamental Research, \\
Homi Bhabha Road,
Bombay 400 005, India}
\date{\today}
\maketitle
\begin{abstract}
We show explicitly that there is particle creation
in a static spacetime. This is
done by studying the field in a coordinate system
based on a physical principle which has
recently been proposed.
There the field is quantized  by decomposing it into
positive and negative frequency modes on a
particular spacelike surface. This decomposition
depends explicitly on the surface where the
decomposition is performed, so that an observer who
travels from one surface to another will
observe particle production due to the different
vacuum state.
\end{abstract}


\section{Introduction}
 Before one can quantize a free field propagating
on a curved background one must have a unique means
of splitting the field into positive and negative
frequency modes. This procedure requires
a unique definition of time with which to perform
the decomposition.

There have been many attempts to define the vacuum
for a free field propagating in a nontrivial spacetime.
It is a
common feature of all these attempts that the choice of
vacuum is determined by a particular choice
of time coordinate. This is true even for such
general quantization procedures as Deutsch and
Najmi \cite{deunaj} although there the dependence
is not explicit. Instead they require a foliation of
 spacetime
by a family of spacelike hypersurfaces which, in
essence, defines ``instants of time" and the normals
to these surfaces define the ``direction of time".
The choice of time
coordinate, in most computations, has usually been  based
on calculational convenience and not
on a
local physical principle.  The physical principle
which we use here is that suggested by Capri and Roy
\cite{Caproy92}. The principle is that on the surface of
instantaneity a 1+1 dimensional Poincar\'e algebra
including the Killing equations for the generators
should be valid.

In a globally hyperbolic spacetime with one
timelike and one spacelike dimension the surface of
instantaneity, in this case a line, for a given
observer is given by the particular spacelike
geodesic which passes through the  point
at which the observer is located
and is normal to the observer's timelike worldline.
The direction of time on this surface is then
defined to be everywhere normal to this spacelike
geodesic. It has been shown that this definition of
time is the unique one to obey the physical
principle given above \cite{Caproy92}.

The coordinates on the spacelike surface are chosen,
for convenience,
to be Riemann coordinates based at the observer's
position, although any other coordinatization of the
spacelike surface will do. To define the direction of
 time Gaussian
Geodesic Normal coordinates are then constructed on
this spacelike surface so that the time coordinate
of some point off the surface is just given by the
distance from the point to the spacelike surface.

When one expresses the metric in terms of these
new coordinates one finds that the metric has
the form,
\begin{equation}
ds^2 = dt^2 + g_{11} dx^2,
\label{0.1}
\end{equation}
with $g_{11} < 0 $ and $g_{11} = -1 + O(t^2) $ near
the origin of coordinates.

To now decompose the field into positive and
negative frequency modes we impose initial
conditions that force the field to have the correct
time dependence $(\exp(-i\omega t))$ in the
neighbourhood of this spacelike surface. To ensure
this correct time dependence we impose the initial
conditions,
\begin{equation}
\phi_n^+(t,x)\left|_{t=0}\right. = A_n(0,x) \ \
{\rm and} \ \ (\partial_t
\phi_n^+(t,x))\left|_{t=0}\right. = -i
\omega_n(0) A_n(0,x).
\label{0.11}
\end{equation}
where the $A_n(0,x)$ are the spatial eigenmodes of the
Laplace-Beltrami operator on the surface $t=0$
where the decomposition is to be performed.

In section II we construct the metric in terms
of these physically preferred coordinates and write
out explicitly the boundary conditions which
determine the positive frequency modes at the
surface $t=0$. In section III a complete set of
modes for the Laplace-Beltrami operator for a
massive scalar field is obtained and in
section IV the orthogonality relation satisfied by
these modes is calculated. In section V we
then use these orthogonality relations to
impose the physically relevant boundary conditions
which were calculated in section II. In section VI
the actual particle creation due to
the presence of the gravitational field is
calculated for a
stationary observer with respect to this spacetime,
in terms of the original static coordinates.

\section{The preferred coordinates}
The spacetime we are interested in is
described by the metric \cite{witten}
\begin{equation}
ds^2=\alpha(X)dT^2-{{dX^2}\over{\alpha(X)}}
\label{1.1}
\end{equation}
where
\begin{equation}
\alpha (X)= 1- \exp({-q(|X|-r)}).
\label{1.2}
\end{equation}
To simplify the technical discussion later on we
choose $r$ such that $\exp(qr)<2$, then
$ -1< \alpha (X) < 1 $.
This spacetime was first investigated by Witten
 \cite{witten} as a $1+1$ dimensional eternal
  blackhole spacetime.

To construct the preferred coordinates we must
 solve the geodesic equations for this
 spacetime. The first integrals of the geodesic
  equations yield
\begin{equation}
\frac{dT}{ds}= \frac{C_0}{\alpha (X)}\ \ \ ; \ \
 \frac{dX}{ds}=
\epsilon_1(C_0^2-\epsilon \alpha
(X))^{\frac{1}{2}}
\label{1.2.1}
\end{equation}
where $\epsilon_1 =\pm 1$ and $\epsilon = -1$
for spacelike geodesics and $\epsilon =1$ for
 timelike geodesics. The spacelike geodesic
 which is perpendicular to the timelike vector
 $\frac{1}{\sqrt{\alpha_0}}(1,0)$,$(\alpha_0\equiv
  \alpha(X_0))$ and which can
 be treated as the tangent vector to the
 worldline of the observer at $P_0 (T_0,X_0)$,
 is given by setting $C_0=0$ and $\epsilon =-1$.
  We can therefore see on
 this surface $S_0$ that $T_0=T_1$. $T_1$ is the
 $T$ coordinate of the point $P_1 (T_1,X_1)$
 which is the point at which the geodesic from the
 general point $P(T,X)$ intersects this spacelike
 surface orthogonally.
  The preferred time coordinate $t$ is given
 by the distance along the timelike geodesic
connecting $P_1$ to the general point $P(T,X)$
which is normal to the surface $S_0$ at $P_1$.
 This timelike geodesic is given by (\ref{1.2.1})
with $C_0^2= \alpha_1\equiv\alpha (X_1)$ and $\epsilon = 1$
so that
\begin{equation}
t = \int_{X_1}^X dX' \frac{ds}{dX'}= \int_{X_1}^X
dX'\frac{\epsilon_1}
{\sqrt{\alpha(X_1)-\alpha(X')}}.
\label{1.2.2}
\end{equation}
One can also calculate the change in the
coordinate $T$ along this geodesic,
\begin{equation}
T - T_1 = T - T_0 = \int _{P_1}^{P} dT =
\int_{X_1}^X dX'\frac{\epsilon_1 \epsilon_2
\sqrt{\alpha(X_1)}}{\alpha(X')
\sqrt{\alpha(X_1)-\alpha(X')}}.
\label{1.2.3}
\end{equation}
These two equations allow us to express the
 metric in terms of the coordinates $t$ and
$X_1$. The preferred coordinate $x$ on $S_0$
is now constructed using a 2-bein of orthogonal
 basis vectors at $P_0$, $e_0(P_0)$ and
 $e_1(P_0)$. With $e_0(P_0)$ given by
  $\frac{1}{\sqrt{\alpha_0}}(1,0)$ the tangent
  to the observer's worldline and $p^{\mu}$ given
   by the tangent vector at $P_0$ to the geodesic
   connecting $P_0$ to $P_1$, the Riemann normal
   coordinates $\eta^{\alpha}$ of $P_1$ are given by
\begin{equation}
s p^{\mu} = \eta^{\alpha}e_{\alpha}^{\mu}(P_0)
\label{1.2.4}
\end{equation}
where $s$ is the distance along the geodesic $P_0-P_1$.
 Using $e_{\alpha}^{\mu}e_{\beta \mu} =
 \eta_{\alpha \beta}$ (Minkowski metric),
  and the orthogonality of $p^{\mu}$ to $e_0(P_0)$
  we have
\begin{equation}
\eta^0=sp^{\mu}e^0_{\mu}(P_0) \ \ \ \ \
 \eta^i=-sp^{\mu}e^i_{\mu}(P_0).
\label{1.2.5}
\end{equation}
The surface $S_0$ is just the surface $\eta^0=0$
and the coordinate $x$ is
\begin{equation}
x=\eta^1 = -sp^{\mu}e^1_{\mu}(P_0)=\int^{X_1}_{X_0}dX'
 \frac{1}{\sqrt{\alpha(X')}}
\label{1.2.6}
\end{equation}

The preferred coordinates $(t,x)$
are then given by solving the above integrals for $X > 0 $
or $X<0$. After choosing $\epsilon_1=-1$ one obtains for $X>0$,
\begin{eqnarray}
T &=&{-2\,{\sqrt{-1 + {e^{q\,\left( -r + { X_1} \right) }}}}\,
      {\rm tan}^{-1} ({\sqrt{-1 + {e^{q\,\left( -X + { X_1} \right)
       }}}})\,{ }\,{ \epsilon_2} \over q}\nonumber \\
&+&{{ 2\,{\rm tanh^{-1}}({{{\sqrt{-1 + {e^{q\,\left( -X + { X_1}
\right) }}}}}\over
         {{\sqrt{-1 + {e^{q\,\left( -r + { X_1} \right) }}}}}})\,{
	 }\,{ \epsilon_2}}\over q} + \,{ T_0}
\label{1.3}
\end{eqnarray}
\begin{equation}
x= {2\over{q}} \left\{ {\rm tanh}^{-1}
(\sqrt{\alpha_1}) -  {\rm tanh}^{-1} (\sqrt{\alpha_0}) \right\}
\label{1.4}
\end{equation}
\begin{equation}
t = {{2\,{e^{{{q\,\left( -r + { X_1} \right) }\over 2}}}\,
     {\rm tan}^{-1} ({\sqrt{-1 + {e^{q\,\left( -X + { X_1} \right)
      }}}})\,{
     }}\over q}
\label{1.5}
\end{equation}
and for $X<0$,
\begin{eqnarray}
T &=&{{2\,{\sqrt{-1 + {e^{-\left( q\,\left( r + { X_1} \right)
 \right) }}}}\,
      {\rm tan^{-1}}({\sqrt{-1 + {e^{q\,\left( X - { X_1} \right)
       }}}})\,{ }\,{ \epsilon_2}}\over q} \nonumber \\
&-& {{2\,{\rm tanh^{-1}}({{{\sqrt{-1 + {e^{q\,\left( X - { X_1}
 \right) }}}}}\over {{\sqrt{-1 + {e^{-\left( q\,\left( r + { X_1}
  \right)  \right) }}}}}})\,{ }\,{ \epsilon_2}}
     \over q} + { T_0}
\label{1.35}
\end{eqnarray}

\begin{equation}
x= {2\over{q}} \left\{ -{\rm tanh}^{-1}
(\sqrt{\alpha_1}) +  {\rm tanh}^{-1} (\sqrt{\alpha_0}) \right\}
\label{1.45}
\end{equation}
\begin{equation}
t = {{-2\,{\rm tan}^{-1} ({\sqrt{-1 + {e^{q\,\left( X - { X_1} \right)
 }}}})\,{ }}\over
   {{e^{{{q\,\left( r + { X_1} \right) }\over 2}}}\,q}}
\label{1.55}
\end{equation}
where
\begin{equation}
 \epsilon_2 = \pm 1.
\label{1.56}
\end{equation}
{}From these coordinate transformations we can see that $x$
and $t$ both
run from $-\infty$ to $+\infty$ and cover the region of the
 original space
corresponding to $\left|X\right| > r$, the region outside the
 horizon. The region inside the horizon is shrunk to a point.
Furthermore, the region between the observer and the horizon
 $(r<|X|<|X_0|)$ is covered twice.

In terms of the coordinates $(t,x)$ the metric is
now
\begin{equation}
ds^2 = dt^2 -\left(1+tp(x)
\tan\left[tp(x)\right]\right)^2dx^2
\label{1.6}
\end{equation}
where
\begin{equation}
p(x) = {q\over{2}} {\rm sech}\left[B(x)\right]
\label{1.7}
\end{equation}
and
\begin{equation}
B(x) = \tanh^{-1} \left[ \sqrt{\alpha_0} \right] +{x q
\over{2}}
\label{1.8}
\end{equation}
We can see at this point that in this coordinate
system, which does have a physical basis, the
metric no longer appears static.

In these new coordinates the Klein-Gordon
equation for a massive scalar field is
\begin{equation}
\partial_t^2 \phi +
\frac{1}{2} \left( \partial_t ln(|g|) \right)\partial_t
\phi +
\frac{1}{\sqrt{|g|}} \partial_x \left(
\sqrt{|g|}g^{11}\partial_x \right) \phi +
m^2 \phi = 0.
\label{1.9}
\end{equation}

We now define instantaneous eigenfunctions $
A_n(t,x)$ of the spatial part of the
Laplace-Beltrami operator, such that
\begin{equation}
\left[\frac{1}{\sqrt{|g|}} \partial_x \left(
\sqrt{|g|}g^{11}\partial_x\right) + m^2  \right]
A_k(t,x) = \omega_k^2(t) A_k(t,x)
\label{1.10}
\end{equation}

The positive frequency solutions of (\ref{1.9}) are
then defined as those which satisfy the initial
conditions
\begin{equation}
\phi_k^+(t,x)\left|_{t=0}\right. = A_k(0,x) \ \
{\rm and} \ \ (\partial_t
\phi_k^+(t,x))\left|_{t=0}\right. = -i
\omega_k(0) A_k(0,x).
\label{1.11}
\end{equation}
These initial conditions ensure that the positive
frequency part of the field has the desired time
dependence at $t=0$. These positive frequency solutions
form a vector space which is made into a Hilbert
space using the standard Klein-Gordon inner product.

{}From the simple form of the metric at $t=0$ we
see that
\begin{equation}
A_k(0,x) = {\rm sin}(2\frac{k}{q}B(x)) \ {\rm or} \ {\rm cos}
(2\frac{k}{q}B(x))\  \ \ {\rm and} \ \ \
\omega_k^2(0) = (k^2 + m^2 )
\label{1.12}
\end{equation}

With the positive frequency solutions defined in
this way we can then write out the quantized field
as
\begin{equation}
\Psi_1=\int_0^{\infty} dk \frac{1}{\sqrt{2 \omega_{k}}}\left\{
\phi_{k}^+(t,x)a_{k} + \phi_{k}^{(+)\ast}(t,x)a_{k}^\dagger
\right\},
\label{1.13}
\end{equation}
where the subscript $1$ of the field simply denotes the
surface on which the positive frequency modes have
been defined.
In this expression we have written
$\omega_k(0)$ as $\omega_k$ and we will continue
this practice.
 Unfortunately  (\ref{1.6})
is too complicated to obtain the general form of
the modes in terms of the coordinates $(t,x)$. This
is, however, not really a problem as the point of
this approach is to find out what boundary
conditions should be imposed. It is therefore
sufficient to solve the field equations in
whatever coordinate system is convenient and then
express these solutions in the preferred coordinate
system to impose the boundary conditions.

\section{Modes of Field Equation}
 From the form of (\ref{1.1}) we can see that in
terms of the original coordinates $(T,X)$ the field
equations are separable. For this reason we
solve for the modes in these coordinates and
then express the solutions in terms of the
preferred
coordinates using the coordinate transformations
given above (\ref{1.3}-\ref{1.55}). In terms of the
coordinates $(T,X)$ the Klein-Gordon operator
has the form
\begin{equation}
\frac{1}{\alpha (X)} \partial_T^2 \phi - (\partial_X
\alpha (X))\partial_X \phi
-\alpha (X) \partial_X^2 \phi +m^2 \phi = 0
\label{2.1}
\end{equation}
By assuming a $T$ dependence for the
field of the form $\exp(-i \omega_p T)$ we obtain the
following differential equation,
\begin{equation}
 \partial_X(
\alpha (X) \partial_X \phi) +({\omega_p^2\over
\alpha(X) }-m^2) \phi = 0
\label{2.2}
\end{equation}

To construct a self-adjoint extension
for this operator we are required to construct solutions
which vanish at the horizon where $\alpha(X)=0$.

By making a change of variable to $ z= 1- \exp(-
q(|X|-r))=\alpha (X) $ we obtain the
following differential equation in terms of $z$,
\begin{equation}
z(1-z)^2 \Psi''(z) + (1-z)(1-2z)\Psi'(z) +
(\frac{p^2}{z}-\mu^2)\Psi(z)= 0.
\label{2.2.10}
\end{equation}
where
\begin{equation}
p^2 = \frac{\omega_p^2}{q^2}\ \ \ {\rm and }\ \ \
\mu^2 = \frac{m^2}{q^2}
\label{2.3}
\end{equation}

We are interested in constructing solutions outside
the horizon so that $ \left|X\right| > r$ and $z > 0$.
As mentioned above we also require that the solutions
vanish at $z=0$.

The two independent
solutions to  (\ref{2.2}) are
\begin{equation}
\Psi_{1p} (z) = z^n (1-z)^l F(a,b,c,z)
\label{2.6}
\end{equation}
where $F(a,b,c,z)$ is an hypergeometric function and
\begin{eqnarray}
n &=& ip \nonumber   \\
l &=& i\sqrt{p^2-\mu^2} \nonumber   \\
a &=& n+l  \nonumber\\
b &=& n+l+1 \nonumber \\
c &=& 1+2n
\end{eqnarray}
and
\begin{equation}
\Psi_{2p} (z) = z^n (1-z)^l F(a,b,c,z)
\label{2.8}
\end{equation}
where
\begin{eqnarray}
n &=& -ip   \nonumber \\
l &=& -i\sqrt{p^2-\mu^2}  \nonumber  \\
a &=& n+l  \nonumber\\
b &=& n+l+1 \nonumber \\
c &=& 1+2n.
\end{eqnarray}

We can now finally write out the desired
solution to (\ref{2.2.10})
\begin{equation}
\Psi(p,X)= \left[\Psi_{2p}(0)
\Psi_{1p}(z)-\Psi_{1p}(0)\Psi_{2p}(z)\right] \epsilon(X).
\label{2.9.2}
\end{equation}

The general solution to (\ref{2.1}) can then be written,
\begin{equation}
\Psi (T,X) = \int_{\mu}^{\infty} dp \left\{
\left(A(p) \Psi(p,X)\exp(-i\omega_pT)
 + B(p) \Psi(p,X)\exp(i\omega_pT)
\right)\right\}
 \label{2.10}
\end{equation}

We now  impose the initial conditions
 (\ref{1.12}) which then give some physical
meaning to the decomposition of this field into
positive and negative frequency parts. To
explicitly impose these initial conditions it is
first  useful to find the orthogonality relation
satisfied by the modes $\Psi(p,X)$.

\section{Orthogonality of Modes}
To find the orthogonality relation satisfied by
the mode  $\Psi(p,X)$
we follow standard Sturm-Liouville
approach and recall  that the mode satisfies,
\begin{equation}
z(1-z)^2 F''(p,z) + (1-z)(1-2z)F'(p,z) +
(\frac{p^2}{z}-\mu^2)F(p,z)=0.
\label{3.1}
\end{equation}
We can also write out a similar equation which is
satisfied by the modes $F^{\ast}(k,z)$. If one now
multiplies the equation for $F(p,z)$ by
$F^{\ast}(k,z)$ and the equation for
$F^{\ast}(k,z)$ by $F(p,z)$ and looks at the
difference of the two equations one can see that
after integrating over the range $z=0$ to $z=1$
and integrating the two terms by parts once we are
left with the relation,
\begin{eqnarray}
\int_{0}^1 dz\frac{F(p,z)F^{\ast}(k,z)}{z(1-z)}&=&
\lim_{z \rightarrow 1 }\frac{z(1- z)}{(p^2-k^2)}
\left(F'(p,z)F^{\ast}(k,z)-
F(p,z)F'^{\ast}(k,z)\right)
  \\
&-&\frac{z(1- z)}{(p^2-k^2)}
\left(F'(p,z)F^{\ast}(k,z)-
F(p,z)F'^{\ast}(k,z)\right)|_{z=0}
\label{3.2}
\end{eqnarray}
Because of the boundary conditions satisfied by
$F(p,z)$ and $F^{\ast}(k,z)$ the second term in
this relation is identically zero. The first term,
as we show, is proportional to a delta
function. This shows that these modes are orthogonal.
To see that this expression is indeed proportional to a
delta function we first smear it with a
smooth function of $p$
and show that the result is
proportional to that function
evaluated at $p=k$. When one attempts to
evaluate the limit in the first term one finds that
all the various terms are proportional to a common
factor which produces the delta function, this
factor is
\begin{equation}
\lim_{z \rightarrow 1} \frac{(1-z)^{-i(\sqrt{k^2-
\mu^2}-\sqrt{p^2-\mu^2})}-(1-z)^{i(\sqrt{k^2-
\mu^2}-\sqrt{p^2-\mu^2})}}{(p-k)}
\label{3.3}
\end{equation}
 To proceed we introduce a
regularization $(1-z)^{\epsilon}$ and write,
\begin{equation}
F(p,z) = \lim_{\epsilon \rightarrow 0}F(p,z)
(1-z)^{ \epsilon}.
\label{3.3.5}
\end{equation}
The integral we must evaluate is,
\begin{equation}
\lim_{\epsilon \rightarrow 0}
\int_{-\infty}^{\infty}dp \frac{f(p)}{p-k}
\lim_{z \rightarrow 1}
(1-z)^{2 \epsilon}\left\{ (1-z)^{-i(\sqrt{k^2-
\mu^2}-\sqrt{p^2-\mu^2})}-(1-z)^{i(\sqrt{k^2-
\mu^2}-\sqrt{p^2-\mu^2})}\right\}
\label{3.3.6}
\end{equation}
This shows that there is no contribution to the
integral from the regions where $|p-k| > R$. In these
regions the pole at $p=k$ is not realized so one may
interchange the order in which the limits are performed.
These contributions then go to zero as the limit $ z
\rightarrow 1$ is performed. We are then left with
the integral
\begin{equation}
\lim_{\epsilon \rightarrow 0}\lim_{R \rightarrow 0}
\int_{k-R}^{k+R}dp \frac{f(p)}{p-k}
\lim_{z \rightarrow 1}
(1-z)^{2 \epsilon}\left\{ (1-z)^{-i(\sqrt{k^2-
\mu^2}-\sqrt{p^2-\mu^2})}-(1-z)^{i(\sqrt{k^2-
\mu^2}-\sqrt{p^2-\mu^2})}\right\}
\label{3.3.7}
\end{equation}
It is now convenient to make a change of variable
to the variable $x$ where
\begin{equation}
x= \frac{k \ln(1-z) (p-k) }{\sqrt{k^2-\mu^2}}
\label{3.4}
\end{equation}
We next expand the integrand in powers of $(p-
k)$ and find that as $R \rightarrow 0$ we are
left with the smooth function $f(p)$ evaluated at
the pole multiplied by a function of $k$.

Using the above analysis for the orthogonality relations
in $z$ one can then write the orthogonality relations in
$X$,
\begin{eqnarray}
\int_{\left|X\right|>r} dX
\frac{\Psi(p,X)\Psi^{\ast}(k,X)}{\alpha(X)}&=&
 \delta (p-k) \left|(A \Psi_{1k}(0) + B
\Psi_{2k}(0) )\right|^2  \nonumber \\
&\equiv& \delta (p-k) \left| F(k) \right|^2  \label{3.5} \\
\nonumber
\end{eqnarray}
where
\begin{eqnarray}
A=& & \frac{\sqrt{2}(k-\sqrt{k^2-\mu^2})\Gamma(1-
2ik)}{\pi\sqrt{k q \sinh(2\pi\sqrt{k^2-
\mu^2}})}\Gamma^2(-i(k-\sqrt{k^2-
\mu^2}))  \nonumber \\
& &\times\sinh(\pi(k-\sqrt{k^2-\mu^2}))
\sinh(\pi(k+\sqrt{k^2-\mu^2}))
\label{3.6}
\end{eqnarray}
\begin{eqnarray}
B=& & \frac{\sqrt{2}(k+\sqrt{k^2-
\mu^2})\Gamma(1+2ik)}{\pi\sqrt{k q
\sinh(2\pi\sqrt{k^2-\mu^2}})}\Gamma^2(-
i(k+\sqrt{k^2-\mu^2})) \nonumber   \\
& &\times\sinh(\pi(k-\sqrt{k^2-
\mu^2})) \sinh(\pi(k+\sqrt{k^2-\mu^2})).
\label{3.7}
\end{eqnarray}

\section{Frequency decomposition and The Vacuum}
We now decompose the field into positive and
negative frequency parts by picking out the
positive frequency part of the field as that which
satisfies the initial conditions in the preferred
coordinates. This allows us to extract the
annihilation operator for this field and thus
define the vacuum state for this field on this
particular spacelike hypersurface (i.e. the
appropriate hypersurface which passes through the
point $(T_0,X_0)$). The physically relevant
question is, of course, how this decomposition
depends on the point $(T_0,X_0)$ which could
represent the position of an observer. If this
decomposition depends on the position of the
observer then at some different position presumably
the observer would observe some sort of particle
density due to the change in composition of the
vacuum state. To see this we must impose the
initial conditions relevant to the quantization on
this surface. Recall the initial conditions
\begin{equation}
\phi_{k}^+(t,x)\left|_{t=0}\right.= A_{k}(0,x)\ \ {\rm
and } \ \
(\partial_t\phi_k^+(t,x))\left|_{t=0}\right.= -
i\omega_k(0) A_k(0,x)
\label{4.1}
\end{equation}
where
\begin{equation}
A_{k}(0,x)= {\rm sin}(2\frac{k}{q}B(x))  \ \ {\rm
and } \ \ \omega_k(0)=(k^2 + m^2 )^{\frac{1}{2}}
\label{4.2}
\end{equation}
We can thus write the general form of the solution
which satisfies these initial conditions for
this particular mode $k$
\begin{equation}
\Psi_k (T,X) = \int_{\mu}^{\infty} dp  \left\{
\left(A_{k}(p) \Psi(p,X)\exp(-i\omega_pT) +
 B_{k}(p) \Psi(p,X) \exp(i\omega_pT)
\right) \right\}
\label{4.3}
\end{equation}
where we now regard $T$,$X$ and $z$ as  functions of
$(t,x)$. This can be easily done given the
coordinate transformations of section II. Because
the initial conditions are imposed at $t=0$ we
need only be concerned with the form of this field
and its derivative normal to $t=0$ for $z(t=0,x)$
 in order to evaluate the expansion
coefficients $A(p,n)$,$A^{\ast}(p,n)$,$B(p,n)$ and
$B^{\ast}(p,n)$.

By using the orthogonality relations calculated in
the last section we can write out the initial
condition equations,

\begin{eqnarray}
\left| F(k) \right|^2
\left(  A_{k}(k) \exp{(-i\omega_{k}T_0)}\right.&+
&\left. B_{k}(k)
\exp{(i\omega_{k}T_0)}\right) \nonumber  \\
&=&\int_{\left|X\right|>r} dX
\frac{\Psi^{\ast}(k,X) {\rm sin}(2\frac{k}{q}B(x))}{\alpha(X)}
\label{4.5}
\end{eqnarray}

\begin{eqnarray}
\left| F(k) \right|^2\left( B_k(k)
\exp{(-i\omega_{k}T_0)}\right.&-&\left. B_{k}(k)
\exp{(i\omega_{k}T_0)}\right)  \nonumber \\
&=&\int_{\left|X\right|>r} dX \frac{\Psi^{\ast}(k,X)
 {\rm sin}(2\frac{k}{q}B(x))}{\alpha(X)(\frac{\partial T}
 {\partial t})
 \left|_{t=0}\right.}
\label{4.5.2}
\end{eqnarray}

In taking the time derivative of (\ref{4.3}) one
does not pick up a $\frac{\partial X}{\partial t}$
 because at $t=0$
this is zero. Again in these expressions it can be
seen that we are still regarding $z$ as $z(0,x)$ and
$x$ is the inverse of this function in the integral.
We have now determined $A(p,n)$,$A^{\ast}(p,n)$,
$B(p,n)$ and $B^{\ast}(p,n)$
and can therefore decompose the field explicitly in
terms of positive and negative frequency modes on
this surface,
\begin{equation}
\Psi_1=\int_0^\infty dk
 \frac{1}{\sqrt{2 \omega_k}}\left\{
\phi_{k1}^+(t,x)a_{1}(k) +
\phi_{k1}^{(+)\ast}(t,x)a_{1}(k)^\dagger \right\}
\label{4.6}
\end{equation}
where the extra subscripts denote the surface on
which the frequency decomposition has been performed
and the modes $ \phi_{k1}^{(+)\ast}(t,x) $ and
$\phi_{k1}^+(t,x)$ are the ones constructed with
the expansion coefficients which satisfy
(\ref{4.5}) and (\ref{4.5.2}).
One may now define the
vacuum relevant to this field on the surface
$(t=0)$ in the usual way,
\begin{equation}
a_{1}(k) \left| 0_1 \right> = 0 \ \ \ \forall\ \ k
\label{4.7}
\end{equation}
where again the subscript denotes ``when" this is the
vacuum state for the field. To see whether
particles are created by the gravitational field in
this spacetime one must look closely at how this
state depends on the surface chosen. So at this stage all
one needs is the normal derivatives, with respect to the
 spacelike surface, of the field on the surface. In Section VI
 the full transformation equations will be required.

\section{Particle Creation}
  To see whether particles are created by the
gravitational field in the spacetime one must look
closely at how  the field
decomposition depends on the surface chosen (i.e.
the position of the observer). To obtain the
spectrum of particles created one must calculate
the Bogolubov transformation between the different
annihilation and creation operators and look at the mixing of
positive and negative frequency parts. To calculate
the Bogolubov transformation we can just match the field
from two different quantizations on a common surface.
The easiest way to do this is to propagate one field to
the surface on which the second is quantized. We can
therefore write
\begin{equation}
\Psi_1(t,x) = \Psi_2(0,x')\ \ \ {\rm and}\ \
 \partial_{t'}\Psi_1(t,x) = \partial_{t'}(\Psi_2(t',x'))
 \left|_{t'=0}\right.
\label{5.1}
\end{equation}
where $t$ is the proper distance between the two
quantization surfaces. This distance will, in general, depend
on where on the surface one calculates the distance.
To make the calculation simpler we
take $X'_0=X_0$ so that the observer is stationary with
respect to the original coordinates where the metric is
static.

Because of the simple form of the modes
at $t=0$ one can calculate the Bogolubov coefficients
and write an expression of the form
\begin{equation}
a_2(k) =\int dp \left( \alpha (p,k) a_1(p) + \beta (p,k)
a_1(p)^{\dagger} \right)
\label{5.2}
\end{equation}
The particle density experienced by an observer travelling
from surface $1$ to surface $2$ is then given by
\begin{equation}
\left| \beta (p,k) \right|^2
\label{5.3}
\end{equation}
In $1+1$ dimensions $\beta(p,k)$ in general has the
form,
\begin{equation}
\beta (p,k) = \int_{-\infty}^{\infty}dx'\frac{q}{2}
\frac{A_{p}(0,x')}{iq\pi\sqrt{\omega_p\omega_k}}
\left\{
\dot{\phi}_{1k}^{+ \ast}(t,x){\partial t\over
\partial t'} + \left(\partial_x\phi_{1k}^{+ \ast}(t,x)\right)
{\partial x\over
\partial t'}
-i\omega_p \phi_{1k}^{+ \ast }(t,x)
 \right\}_{t'=0}.
\label{5.4}
\end{equation}
In this equation the factors  ${\partial t\over \partial t'}$
and ${\partial x\over \partial t'}$ are required because we
are matching the field's  normal derivative
is done with respect to the second surface.

To calculate an approximate form of $\beta$,
valid for short time intervals, we
 expand the integrand
about $t=0$. To
$O(t^2)$ we obtain,
\begin{equation}
\beta(p,k)=-\int_{-\infty}^{\infty}\!\!\! dy \frac{\sin(
\frac{2p}{q}y)
}{q\pi\sqrt{\omega_p
\omega_k}}\tanh^2(y)\sin(\frac{2k}{q}y)p^2(x')\omega_k
(T'_0-T_0)^2
\label{5.5}
\end{equation}
where $p(x')$ is given by (\ref{1.7}) and we have
changed variables from $x'$ to $y=B(x')$. In getting from
(\ref{5.4}) to (\ref{5.5}) the second term of (\ref{5.4}) doesn't
contribute to the integral because it is odd in $y$.  It
 should
be restated that this is particle creation observed by an
 observer stationary with respect to the original static
coordinates.
 This can now be
rewritten as,
\begin{equation}
\beta(p,k)=-\frac{(T'_0-T_0)^2\omega_k q}{4\pi} \sqrt{
\frac{1}{\omega_k\omega_p}}
\int_{-\infty}^{\infty}\!\!\!dy \frac{\sin(\frac{2p}{q}y)}{
{\rm cosh}^2(y)}\tanh^2(y)
\sin(\frac{2k}{q}y).
\label{5.7}
\end{equation}

Several comments are in order here. Firstly, $\beta(p,k)$
is clearly non-zero so that particles are produced
in this short time interval $\delta t=T'_0-T_0$.

Secondly, our approximation clearly only holds for
 $\omega_k < q$ since the expansion breaks down for
 $\omega_k\delta t>1$. This means
that we can only crudely estimate the number of particles
produced in the time $\delta t$ since an ultraviolet cutoff of
$k=\sqrt{q^2-m^2}$ is required.

Putting all this together  we see that the momentum density of
particles labelled by $k$ produced in the time interval $t$ is:
\begin{eqnarray}
n_t(k)&=&\int_0^{\infty} dp \left| \beta(p,k) \right|^2
\nonumber \\
&\simeq & \frac{q^2\delta t^2}{4\pi}\frac{\sqrt{\omega_k}}{q}
\int_0^{\infty} \frac{dp}{\sqrt{p^2+m^2}}\frac{\pi^2}{q^2}\left|
\frac{(p+k)(\frac{(p+k)^2}{q^2}+3)}{\rm{sinh}\frac{\pi(p+k)}{q}}-
\frac{(p-k)(\frac{(p-k)^2}{q^2}+3)}{\rm{sinh}\frac{\pi(p-k)}{q}}
\right|^2
\label{5.8}
\end{eqnarray}

Within the spirit of the approximation, the total number of
particles created in the time $\delta t$ with $\omega_k < q $ is:
\begin{equation}
N_t=\int_0^{\sqrt{q^2-m^2}}n_t(k) dk
\label{5.9}
\end{equation}
This integral is finite, of course. If the upper limit is
allowed to
go to $\infty$ then the integral diverges linearly. This does
not mean that the Bogolubov transformation is not unitarily
implementable. Our approximations simply break down and our results
are inconclusive. The difficulty arises from the fact that
there are two time scales namely $T_1=\frac{1}{q}$ and
$T_2=\frac{1}{\omega_k}$. For a fixed $\omega_k$ it is possible to expand
in $\frac{\delta t}{T}$ where $T$ is the smaller of $T_1,T_2$. However,
if $\omega_k$ is unbounded no such expansion is possible.

\section{Conclusions}
  We have shown that although a spacetime may
   be static this
may not preclude particle creation which is a time
dependent phenomenon \cite{labonte} as the gaussian coordinatization
may not be static. The only metrics
which always lead to static Gaussian coordinates
are those which have been
dubbed ``ultrastatic'' by Fulling \cite{fulling}.
We have shown explicitly in
this simple 1+1 dimensional case how the choice of
which coordinates should be used leads to some
interesting results. In particular, the coordinates
which are chosen via a physical principle seem to
suggest that although the spacetime may be manifestly
static
in one coordinate system these may not be the
coordinates that one should use to quantize a field
propagating in the spacetime.

Unfortunately the analysis to find out whether the Bogolubov
transformation is unitarily implementable was inconclusive.
This is due to the fact that the approximate form of $\beta$
which was analysed was not valid for large $k$.

\section{Acknowledgements}
This research was supported in part by a grant from
the Natural Sciences and Engineering Research Council
of Canada. We would also like to thank Stephen Fulling
for pointing out an error in an earlier version of this paper
and also for providing us with other useful insights.


\begin{references}
\bibitem[*]{email} Electronic Mail: lamb@phys.ualberta.ca
\bibitem{deunaj} D. Deutsch and A. Najmi, Phys. Rev.
{\bf D28}, 1907, (1983).
\bibitem{Caproy92} A.Z. Capri and S.M. Roy, Modern
 Physics Letters
A, {\bf 7}, 2317, 1992, also \\
 International Journal of Modern Physics A, {\bf 9}, 1239, (1994).
\bibitem{witten} E. Witten, Phys. Rev. {\bf D44}
(1991) 314. The metric studied in this paper can
be obtained after
some manipulation of the metric in equation (22)
of Witten's paper.
\bibitem{labonte} G. Labont\'{e}, Can. J. Phys.,
 {\bf 53}, 1533, (1975).
\bibitem{fulling} S.A. Fulling,  Aspects of
quantum field theory in curved
spacetime, Cambridge University Press, (1989).
\end{references}
\end{document}